\documentclass[aps,prl,preprintnumbers,superscriptaddress,twocolumn,nofootinbib]{revtex4}
\usepackage[dvips]{graphicx}
\usepackage{bm,latexsym,amsmath,amssymb,amsfonts,mathrsfs}
\usepackage{color}
\input{colordvi.tex}

\begin{document}

\preprint{ CCTP-2018-12, ITCP-IPP 2018/9, KOBE-COSMO-18-08, MAD-TH-18-05}

\title{Weak Gravity Conjecture from Unitarity and Causality}

\author{Yuta Hamada}
\affiliation{Crete Center for Theoretical Physics, Institute for Theoretical and Computational Physics, Department of Physics, University of Crete, P.O. Box 2208, 71003 Heraklion, Greece}

\author{Toshifumi Noumi}
\affiliation{Department of Physics, Kobe University, Kobe 657-8501, Japan}
\affiliation{Department of Physics, University of Wisconsin-Madison, Madison, WI 53706, USA}

\author{Gary Shiu}
\affiliation{Department of Physics, University of Wisconsin-Madison, Madison, WI 53706, USA}

\begin{abstract}

The weak gravity conjecture states that quantum gravity theories have to contain a charged state with a charge-to-mass ratio bigger than unity. By studying unitarity and causality constraints on higher derivative corrections to the charge-to-mass ratio of extremal back holes, we demonstrate that heavy extremal black holes can play the role of the required charged state under several assumptions. In particular, our argument is applicable when the higher spin states Reggeizing graviton exchange are subdominant in the photon scattering. It covers (1) theories with light neutral bosons such as dilaton and moduli, and (2) UV completion where the photon and the graviton are accompanied by different sets of Regge states just like open string theory. Our result provides an existence proof of the weak gravity conjecture in a wide class of theories, including generic string theory setups with the dilaton or other moduli stabilized below the string scale.
\end{abstract}

\maketitle

\section{Introduction}

One of the greatest appeals of string theory is that it provides a consistent framework for constructing a variety of models for particle physics and cosmology while incorporating quantum gravity. The space of consistent string vacua is often known as the string landscape. This existence of a
rich landscape does not however imply that anything goes. 
It has become increasingly clear
that not every seemingly consistent quantum field theory (QFT) models can be consistently embedded into quantum gravity. 
Theories that are not ultraviolet (UV) completable when we turn on gravity are said to live in the swampland~\cite{Vafa:2005ui} (see also~\cite{Brennan:2017rbf} for a review). Thus, identifying nontrivial ultraviolet constraints on QFTs can offer an interesting 
opportunity to probe the nature of quantum gravity phenomenologically.

Among the criteria that distinguish the landscape from the swampland,
the weak gravity conjecture (WGC)~\cite{ArkaniHamed:2006dz} is arguably the most well studied one. Its mild form states that quantum gravity theories have to contain {\it a charged state} with the charge-to-mass ratio $z$ bigger than unity. In $D=4$, this bound is given by
\begin{align}
\label{WGC}
z=\frac{\sqrt{2}M_{\rm Pl}|q|}{m}\geq1\,,
\end{align}
where $M_{\rm Pl}$ is the reduced Planck mass. This conjecture is motivated by black hole (BH) thought experiments and has passed various nontrivial checks in string theory examples~\cite{ArkaniHamed:2006dz}. 
Moreover, arguments based on
holography~\cite{Nakayama:2015hga,Harlow:2015lma,Benjamin:2016fhe,Montero:2016tif}, cosmic censorship~\cite{Horowitz:2016ezu,Cottrell:2016bty,Crisford:2017gsb,Yu:2018eqq}, black holes and entropy consideration~\cite{Cottrell:2016bty,Hebecker:2017uix,Cheung:2018cwt}, dimensional reduction~\cite{Brown:2015iha,Brown:2015lia,Heidenreich:2015nta,Heidenreich:2016aqi,Lee:2018urn} and infrared consistency~\cite{Cheung:2014ega,Andriolo:2018lvp}
have given further evidence for the conjecture. While these recent developments have significantly expanded our view of the WGC, it is fair to say that our understanding is still not complete and further studies toward a proof of the WGC are desired.

The purpose of this Letter is to provide an existence proof of the WGC in certain classes of theories, based on unitarity and causality.
In particular we argue that even if there exists no particle satisfying the WGC bound~\eqref{WGC}, heavy extremal BHs play the role of the required charged state in the following two classes of theories:
\begin{enumerate}
\item Theories with a parity-even light neutral scalar, such as dilaton and moduli, or a spin $s\geq2$ light neutral particle\footnote{This is the same setup considered in \cite{Cheung:2018cwt} to motivate the WGC from an entropy perspective. However, in the Supplemental Material, we point out a loophole in their argument. Also, as we shall see, our unitarity argument can put stronger constraints in this setup and furthermore has wider applicabilities.}. Here ``light" means lighter than the scale $\Lambda_{\rm QFT}$ where the quantum gravity effects come in and the QFT description breaks down.

\item UV completion where the photon and the graviton are accompanied by different sets of Regge states (just like open string theory), and those associated to the graviton are subdominant in the photon scattering.

\end{enumerate}
These two classes cover a wide variety of theories, including generic stringy setups, providing a strong evidence of the mild form of WGC.  We focus on the $D=4$ case in this Letter, and relegate the extension to general spacetime dimension $D\geq5$ to the Supplemental Material.

\section{Strategy}

One might wonder whether our claim is trivial because the extremal charged BHs in the Einstein-Maxwell theory saturates the bound $z=1$. However, it is not true because the BH solutions are modified by higher derivative corrections and so is the charge-to-mass ratio of extremal BHs accordingly~\cite{Kats:2006xp}.

Suppose that the theory is described by the photon and the graviton in the infrared. In $D=4$ their general effective action up to four-derivative operators is then given  by\footnote{Even though we consider a single $U(1)$ for simplicity, generalization to the multiple $U(1)$ case is straightforward. In particular, it trivially follows from our result that there exists heavy extremal BHs with $z>1$ in any charge direction under our assumptions.}
\begin{align}
S&=\int d^4x\sqrt{-g}
\bigg[
\frac{M_{\rm Pl}^2}{2}R-\frac{1}{4}F_{\mu\nu}F^{\mu\nu}
+\frac{\alpha_1}{4M_{\rm Pl}^4}(F_{\mu\nu}F^{\mu\nu})^2
\nonumber
\\
\label{EFT}
&\qquad
+\frac{\alpha_2}{4M_{\rm Pl}^4}(F_{\mu\nu}\widetilde{F}^{\mu\nu})^2
+\frac{\alpha_3}{2M_{\rm Pl}^2}F_{\mu\nu}F_{\rho\sigma}W^{\mu\nu\rho\sigma}
\bigg]
\,,
\end{align}
where $W_{\mu\nu\rho\sigma}$ is the Weyl tensor and $\widetilde{F}_{\mu\nu}=\epsilon_{\mu\nu\rho\sigma}F^{\rho\sigma}/2$. Also we assumed parity invariance for simplicity. In general, we can add parity violating terms like $F_{\mu\nu}F^{\mu\nu}F_{\rho\sigma}\widetilde{F}^{\rho\sigma}$, but they do not change the extremality condition at the leading order.
Note that other four-derivative operators such as $R_{\mu\nu}^2$ are absorbed into the three operators displayed in the above by field redefinition. The higher derivative operators modify black hole solutions, so that the charge-to-mass ratio of extremal black holes are corrected as~\cite{Kats:2006xp}
\begin{align}
\label{correction_to_z}
z=\frac{\sqrt{2}M_{\rm Pl}|Q|}{M}=1+\frac{2}{5}\frac{(4\pi)^2}{Q^2}(2\alpha_1-\alpha_3)\,,
\end{align}
where $M$ and $Q$ are the mass and charge of the black hole, respectively. This formula is applicable as long as the higher derivative corrections are small. More explicitly, it is applicable if the black hole is sufficiently heavy,
\begin{align}
M^2\sim Q^2M_{\rm Pl}^2\gg \alpha_i M_{\rm Pl}^2\,,
\end{align}
because extremal BHs in the Einstein-Maxwell theory satisfy $R\sim M_{\rm Pl}^4/M^2$ and $F^2\sim M_{\rm Pl}^6/M^2$.

An important observation made in~\cite{Kats:2006xp} is that extremal BHs (in the mass range $M^2\gg \alpha_i M_{\rm Pl}^2$) have the charge-to-mass ratio bigger than unity $z\geq1$, {\it if the Wilson coefficients $\alpha_i$ satisfy the condition,}
\begin{align}
\label{bound_on_alpha}
2\alpha_1-\alpha_3\geq0\,.
\end{align}
On the other hand, if $2\alpha_1-\alpha_3<0$, the expectation is no more valid that extremal BHs satisfy the WGC bound. In the rest of this paper we show that the bound~\eqref{bound_on_alpha} with a strict inequality indeed follows from unitarity and causality in the aforementioned two classes of setups.

\begin{figure}
\includegraphics[width=70mm, bb=0 0 579 345]{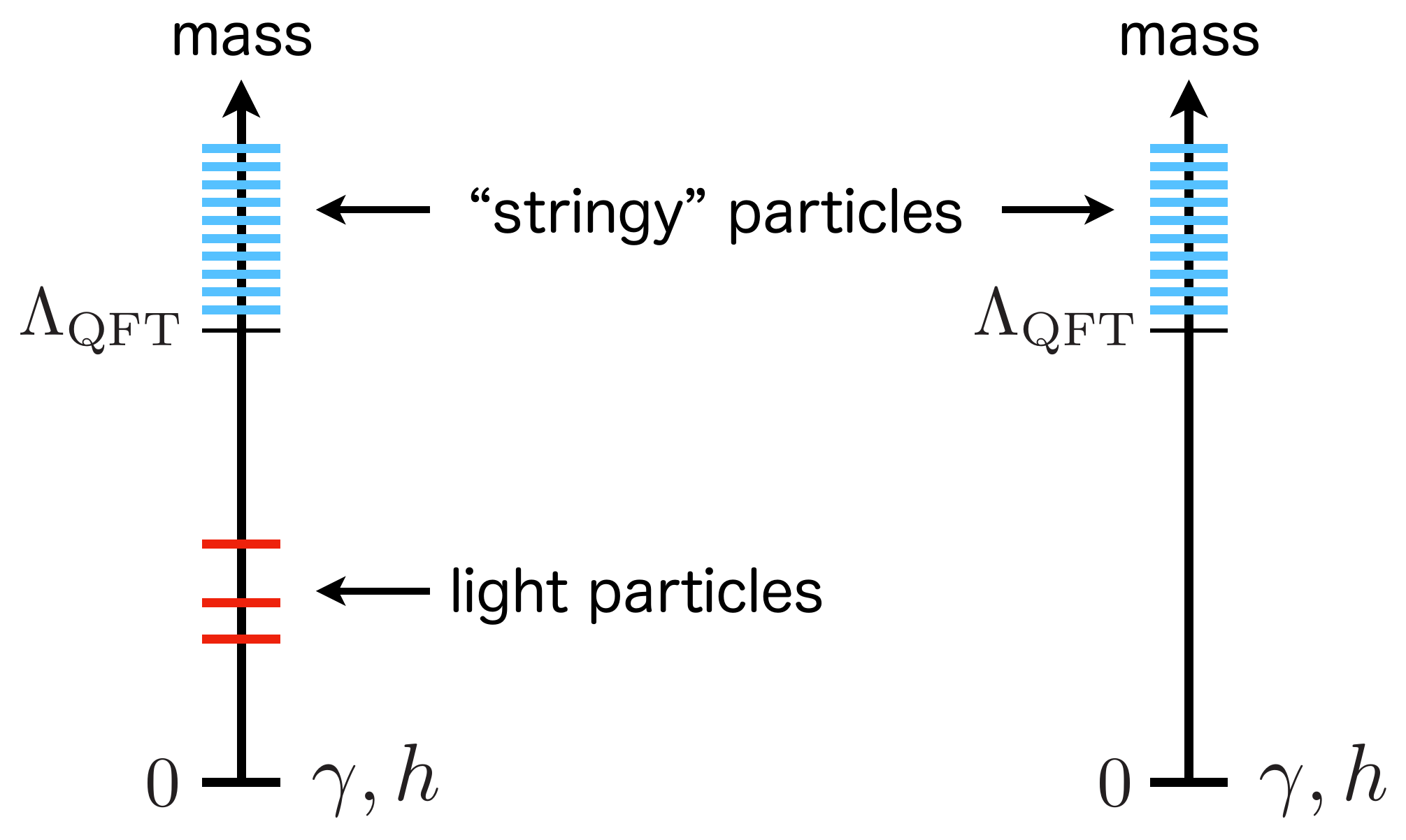}
\caption{A schematic picture of the particle spectrum: We assume that photon and graviton control the BH dynamics in the infrared. The ordinary QFT description breaks down at $\Lambda_{\rm QFT}$, which corresponds to the string scale $M_s$ in string theory. The spectrum may contain light particles below $\Lambda_{\rm QFT}$ (left), but it is also possible that there are no such light particles (right).}
\label{Fig:spectrum}
\end{figure}

\section{Unitarity constraints}

We then summarize the unitarity constraints on the Wilson coefficients $\alpha_i$. For this purpose, let us clarify our setup by classifying possible sources of higher dimensional operators. Fig.~\ref{Fig:spectrum} shows a schematic picture of the particle contents we have in mind. First, we assume that the BH dynamics is controlled by photon and graviton in the infrared, and they are weakly coupled. We also assume a weakly coupled UV completion of gravity throughout the paper. There will be some high energy scale $\Lambda_{\rm QFT}$ where the ordinary QFT description breaks down. Generically, it is below the Planck scale $\Lambda_{\rm QFT}\ll M_{\rm Pl}$. For example, in string theory it is the string scale $\Lambda_{\rm QFT}\sim M_s$, beyond which we have to follow the dynamics of infinitely many local fields and thus the ordinary QFT description breaks down. Note that $M_s\ll M_{\rm Pl}$ in the perturbative string.

Below the scale $\Lambda_{\rm QFT}$, there may exist massive particles, which we call light particles because their masses are smaller than $\Lambda_{\rm QFT}$. Their contributions to higher dimensional operators are qualitatively different 
between the neutral and charged cases
as we explain below.
\begin{enumerate}
\renewcommand{\labelenumi}{(\alph{enumi})}
\setcounter{enumi}{0}
\item Light neutral bosons  (ex. dilaton, axion, moduli)

First, light neutral bosons may generate the effective interactions $\alpha_i$ at the tree-level. Let us consider the dilaton $\phi$ and the axion $a$ for example:
\begin{align}
\label{dilaton}
\mathcal{L}_\phi=
-\frac{1}{2}(\partial_\mu\phi)^2-\frac{m_\phi^2}{2}\phi^2+\frac{\phi}{f_\phi}F_{\mu\nu}F^{\mu\nu}\,,
\\
\label{axion}
\mathcal{L}_a=
-\frac{1}{2}(\partial_\mu a)^2-\frac{m_a^2}{2}a^2+\frac{a}{f_a}F_{\mu\nu}\widetilde{F}^{\mu\nu}\,,
\end{align}
where $m$ and $f$ are the mass and the decay constant, respectively. Integrating out the dilaton and axion, we obtain the tree-level effective couplings,
\begin{align}
\label{dilaton_alpha}
\alpha_1=\frac{2M_{\rm Pl}^4}{m_\phi^2f_{\phi}^2}\,,
\quad
\alpha_2=\frac{2M_{\rm Pl}^4}{m_a^2f_a^2}\,.
\end{align}
More generally, the size of the effective couplings can be estimated as
\begin{align}
|\alpha_i|\gtrsim \mathcal{O}\Big(\frac{M_{\rm Pl}^2}{m_i^2}\Big)\,,
\end{align}
which is indeed the case for the above examples if we assume $f\lesssim M_{\rm Pl}$. Also in the above examples, the signs of the Wilson coefficients are always positive:
\begin{align}
\label{unitarity_alpha}
\alpha_1>0\,,
\quad
\alpha_2>0\,,
\end{align}
which is a consequence of unitarity. More generally, unitarity implies that $\alpha_1>0$ when photon is coupled to a parity-even neutral scalar or a spin $s\geq2$  neutral particle. Similarly, $\alpha_2>0$ when photon is coupled to a neutral pseudo-scalar or a spin $s\geq2$ neutral particle. Note that the spin $s\geq2$ particle may carry an arbitrary parity in either case. See the Supplemental Material for our derivation\footnote{
\label{footnote:positivity}
To our knowledge, there is no explicit derivation of the bound~\eqref{unitarity_alpha} in the literature. The bound was suggested in the seminal work~\cite{Adams:2006sv}, but an explicit derivation of positivity bounds was demonstrated only in a scalar field model (see however~\cite{Bellazzini:2016xrt} for its extension to external spinning particles). Also, in~\cite{Cheung:2014ega}, it was claimed that the bound~\eqref{unitarity_alpha} follows from unitarity by a spectral decomposition argument. However, the interactions that \cite{Cheung:2014ega} can cover are restrictive (allowing only intermediate states with spin $0$ and $2$). Moreover, the interaction for spin $2$ is singular in the UV. We thank Grant Remmen for discussion on this point. In the Supplement Material, we provide a derivation of the bound~\eqref{unitarity_alpha} and clarify under which conditions the bound is applicable especially in the presence of gravity.}.

\begin{figure}
\includegraphics[width=85mm, bb=0 0 337 113]{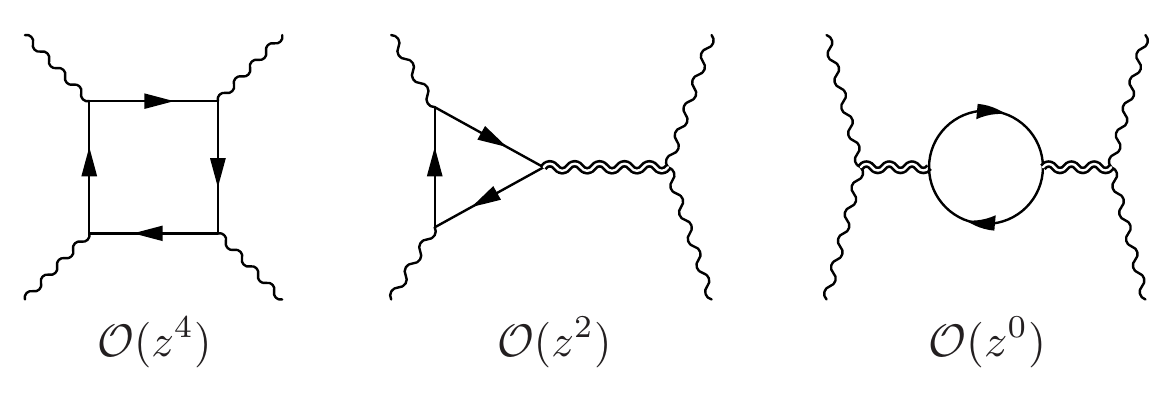}
\caption{Typical one-loop corrections to the $F^4$ terms: In the left figure the massive charged particle (solid line) induces four-point interactions of photon (wavy line) through the gauge coupling, hence it is proportional to $q^4\propto z^4$. In the other two, the diagrams involve graviton (double wavy line). If $z\gg1$, gravity is negligible, so that we may apply the positivity bound derived in non-gravitational theories. The same argument holds more generally, where the charge-to-mass ratio $z$ is replaced by the ratio of the photon coupling of the massive particle and the gravitational force.}
\label{Fig:loop}
\end{figure}

\item Light charged bosons and fermions

In contrast to neutral bosons, charged bosons and fermions cannot generate the effective couplings $\alpha_i$ at the tree-level, hence the leading contribution is at one-loop. For example, the one-loop effective coupling generated by a minimally coupled massive charged particles are estimated as (cf. Fig.~\ref{Fig:loop})\footnote{The running of coupling constants are included in the $\mathcal{O}(1)$ effect, which is valid as long as we are in the perturbative regime.} 
\begin{align}
\alpha_{1,2}=\max\left\{\mathcal{O}(z^4),\mathcal{O}(1)\right\}\,,
\,\,
\alpha_3=\mathcal{O}(z^2)\,,
\end{align}
where $z$ is the charge-to-mass ratio of the particle integrated out. Notice here that when the particle has a large charge-to-mass ratio $z\gg1$, the Wilson coefficients enjoy $|\alpha_1|,|\alpha_2|\gg|\alpha_3|\gg1$. Moreover, $\alpha_1,\alpha_2>0$ follows from unitarity for $z\gg1$, where gravity is negligible compared to the electric force. On the other hand, we have $\alpha_i=\mathcal{O}(1)$ for $z\lesssim 1$. In this regime, as far as we know, no rigorous bound on $\alpha_i$ is known so far essentially because gravity is not negligible.

More generally, when the interaction between photon and the massive particle is stronger than the gravitational force, there exists the hierarchy $|\alpha_1|,|\alpha_2|\gg|\alpha_3|$ and the positivity of $\alpha_1$ and $\alpha_2$ follows from unitarity. If the two interactions are comparable, there is no known rigorous bound, but the induced effective interaction is very small $\alpha_i=\mathcal{O}(1)$ compared to other sources. 
\end{enumerate}
On top of these possible effects of light particles, there are higher derivative corrections from the UV completion of gravity, which we call the UV effects:
\begin{enumerate}
\renewcommand{\labelenumi}{(\alph{enumi})}
\setcounter{enumi}{2}
\item UV effects

From the effective field theory (EFT) point of view, this effect is suppressed by the scale $\Lambda_{\rm QFT}$, where the quantum gravity effects come in and the ordinary QFT description breaks down. Generically, we have\footnote{
One would expect a hierarchy $|\alpha_1|,|\alpha_2|\gg |\alpha_3|$, but it is not a general statement. In this estimate, we assumed that there is a single scale $\Lambda_{\rm QFT}$ and other dimensionless constants are $\mathcal{O}(1)$, which corresponds to assuming $m\sim f$ in the Lagrangian \eqref{dilaton}-\eqref{axion}. The estimate changes, e.g., when $m\ll f\sim M_{\rm Pl}$. As we shall see, another ingredient such as causality or symmetry is necessary to have a hierarchy $|\alpha_1|,|\alpha_2|\gg |\alpha_3|$ in general. Our point here is simply that the Wilson coefficients $\alpha_i$ are suppressed by $\Lambda_{\rm QFT}$.}
\begin{align}
\label{estimate}
\alpha_{1,2}=\mathcal{O}\Big(\frac{M_{\rm Pl}^4}{\Lambda_{\rm QFT}^4}\Big)\,,
\quad\alpha_3=\mathcal{O}\Big(\frac{M_{\rm Pl}^2}{\Lambda_{\rm QFT}^2}\Big)\,,
\end{align}
which corresponds, e.g., to the $\alpha'$ corrections in string theory. In general it is difficult to fix the sign of this effect within the EFT framework without knowing the details of the UV completion of gravity. However, as we discuss in the Supplemental Material, $\alpha_1>0$ and $\alpha_2>0$ follow from unitarity as long as the higher spin states Reggeizing graviton exchange are subdominant in the photon scattering. This may happen, e.g., when the photon and the graviton are accompanied by different sets of Regge states, just as in open string theory.
\end{enumerate}
The magnitude of the three effects (a)-(c) and the unitarity constraints on them are summarized in Table~\ref{table}. In particular, the loop effect (b) may be further classified into two, (b-1) and (b-2), by the size of interactions between the photon and the massive particle.

\begin{table}[t]
\begin{center}\begin{tabular}{l||c|c}
 & magnitude & unitarity \\[0mm]
 \hline\hline(a) neutral bosons & $\displaystyle\alpha_i\gtrsim\mathcal{O}\Big(\frac{M_{\rm Pl}^2}{m^2}\Big)$ & $\alpha_1,\alpha_2>0$   \\[2mm]
 \hline (b) loop effects &&
 \\
\,\,(b-1) $z\gg1$ & $|\alpha_1|,|\alpha_2|\gg|\alpha_3|\gg1$ & $\alpha_1,\alpha_2>0$ 
 \\
\,\,(b-2) $z=\mathcal{O}(1)$ & $\alpha_i=\mathcal{O}(1)$ & N.A. 
 \\\hline (c) UV effects& $\displaystyle
 \begin{array}{c}\displaystyle \alpha_{1,2}=\mathcal{O}\Big(\frac{M_{\rm Pl}^4}{\Lambda_{\rm QFT}^4}\Big) \\ [3mm] \displaystyle \alpha_3=\mathcal{O}\Big(\frac{M_{\rm Pl}^2}{\Lambda_{\rm QFT}^2}\Big)\end{array}$ & 
$\alpha_1,\alpha_2>0\, (\star)$
 \end{tabular} \caption{Sources of higher derivative operators: The tree-level effect (a) from neutral bosons and the loop effect (b-1) give a positive contribution to $\alpha_1$ and $\alpha_2$ (if any) as a consequence of unitarity. The same bounds are applicable to the UV effects (c) if the Regge states associated to the graviton are subdominant in the photon scattering.}
\label{table}
\end{center}
\end{table}

\section{WGC from unitarity}

\begin{figure*}
\includegraphics[width=180mm, bb=0 0 449 112]{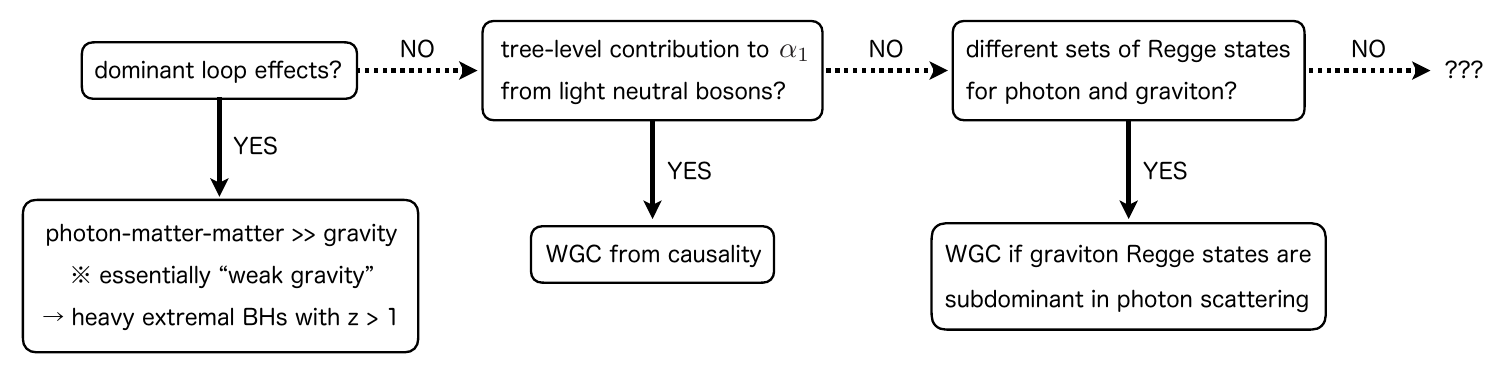}
\caption{A flow chart for our derivation of the WGC from unitarity: Each step explains which conditions are necessary besides unitarity to show that heavy extremal BHs have the charge-to-mass ratio $z>1$ and thus the mild form of WGC is satisfied.}
\label{Fig:flow}
\end{figure*}

We now discuss implications of unitarity on the WGC. See also Fig.~\ref{Fig:flow} for a summary of our argument. One easy observation is that the inequality~\eqref{bound_on_alpha} is satisfied when the effect (b-1) dominates over the others because its contribution to the l.h.s. of Eq.~\eqref{bound_on_alpha} is always positive. This is the case, e.g., when there exists a massive charged particle with $z\gg1$. Since this particle trivially satisfies the WGC bound, this situation is not what we would like to explore\footnote{We note that, even in this situation, extremal BHs can satisfy the WGC in addition to the massive charged particle.}: We are interested in whether extremal BHs may play the role of the charged state required by the WGC in case there are no particles with $z\geq1$. Also, the effect (b-2) is always subleading at least as long as $\Lambda_{\rm QFT}\lesssim M_{\rm Pl}$. Therefore, in nontrivial setups for our question, the loop effect (b) from light particles is always subleading.

Let us then focus on the tree-level effects (a) and (c) in the following. As we explained, $\alpha_1$ and $\alpha_2$ are well constrained by unitarity, but no rigorous bound on $\alpha_3$ is known so far, as far as we know. Since the inequality~\eqref{bound_on_alpha}  involves $\alpha_3$, one might give up deriving it from unitarity. However, it is useful to recall that the $\alpha_3$ operator is significantly constrained by causality.

\subsection{Causality constraints}

The key is that $\alpha_3$ generates new photon-photon-graviton helicity amplitudes which do not exist in the Einstein-Maxwell theory. The photon-photon-graviton amplitudes in the setup~\eqref{EFT} are schematically given by
\begin{align}
\mathcal{M}(1^+,2^-,3^{\pm 2})
=\mathcal{M}(1^-,2^+,3^{\pm 2})
&\sim\frac{E^2}{M_{\rm Pl}}\,,
\nonumber
\\
\mathcal{M}(1^+,2^+,3^{+2})
=\mathcal{M}(1^-,2^-,3^{-2})
&\sim\alpha_3\frac{E^4}{M_{\rm Pl}^3}\,,
\nonumber
\\
\label{helicity_amp}
\text{(other helicity amplitudes)}&=0\,,
\end{align}
where $\mathcal{M}(1^+,2^+,3^{+2})$ stands for the scattering amplitude of two helicity plus photons and one helicity plus graviton (in the all in-coming notation) for example. Also $E$ is a typical energy scale.

In~\cite{Camanho:2014apa}, an interesting observation was made that the new helicity amplitudes lead to causality violation at the energy scale $E\sim M_{\rm Pl}/\alpha_3^{1/2}$, so that this scale has to be beyond the EFT cutoff.
Moreover, it was argued that an infinite tower of massive higher spin particles (just like string theory!) with the lightest particle at the scale  $m\sim M_{\rm Pl}/\alpha_3^{1/2}$ is required to UV complete the EFT at the tree-level without causality violation (see also~\cite{Li:2017lmh,Afkhami-Jeddi:2018own} for a holographic derivation based on the conformal bootstrap approach)\footnote{
Note that QED also generates the effective coupling $\alpha_3$ at one loop (after integrating out the massive electron). In this case, the potential causality problem is fixed by the electron positron pairs along the $t$-channel, without requiring the tree-level exchange of massive higher spin particles~\cite{Camanho:2014apa}. See also~\cite{Goon:2016une} for detailed study of causality in QED.}. In other words, the ordinary QFT description with a finite field content is not available beyond the scale $\sim M_{\rm Pl}/\alpha_3^{1/2}$, hence $\Lambda_{\rm QFT}\lesssim M_{\rm Pl}/\alpha_3^{1/2}$. Therefore, $\alpha_3$ generated at the tree-level is suppressed as\footnote{
To be precise, the bound in $D=4$ accommodates a logarithmic correction as $\alpha_3\lesssim  \frac{M_{\rm Pl}^2\ln \Lambda_{\rm QFT}}{\Lambda_{\rm QFT}^2}$, even though it does not change our conclusion. See~\cite{Camanho:2014apa,Afkhami-Jeddi:2018own} for details.}
\begin{align}
\text{tree-level effects: }\alpha_3 \lesssim  \frac{M_{\rm Pl}^2}{\Lambda_{\rm QFT}^2}\,,
\end{align}
which means that all the tree-level contribution to $\alpha_3$ is classified into the effect (c) in Table~\ref{table} and thus the effect (a) from light neutral bosons has no contribution to $\alpha_3$.

\subsection{Case (1): theories with light neutral bosons}

We now find that if the tree-level effect (a) of light neutral bosons dominates over the others, the Wilson coefficients enjoy
\begin{align}
|\alpha_1|\,,|\alpha_2|\gg|\alpha_3|
\end{align}
as a consequence of causality. Since the effect (a) gives a positive contribution to $\alpha_1$ as a consequence of unitarity, the inequality~\eqref{bound_on_alpha} and thus the mild form of the WGC are satisfied. Recall that we need a parity-even neutral scalar or a spin $s\geq2$ neutral particle to have nonzero $\alpha_1$. We therefore conclude that the mild form of WGC is satisfied by heavy extremal BHs even if there are no charged particles with $z\geq1$, as long as the
photon is coupled to
a parity-even neutral scalar or a spin $s\geq2$  neutral particle with a mass $m\ll \Lambda_{\rm QFT}$.
The dilaton and moduli may play the role of this neutral particle (as long as they are not too heavy), hence this scenario is quite generic.

We also remark that our findings match well with the expectation from
open-closed string duality\footnote{We thank Cumrun Vafa for sharing this observation with us.}.
In string theory, charged particles are generically associated to open strings. If their charge-to-mass ratios do not satisfy the WGC bound $z<1$, the open string has to be long such that its lowest mode is heavy enough to make $z$ small. In this regime, it is more appropriate to interpret the open string loop as a tree-level exchange of closed strings, which naturally gives the tree-level effect (a) from light neutral particles such as dilaton and moduli. 

\subsection{Case (2): open string type UV completion}

Then, what is the case without light neutral bosons? As mentioned, it is possible to give  rigorous bounds on $\alpha_{1,2}$ if the photon and the graviton are accompanied by different sets of Regge states. As an illustrative example, let us consider open string theory: The Regge states associated to the photon and the graviton are the open and closed string states, respectively. Since the open string coupling $g_o$ is parametrically bigger than the closed string coupling $g_s$, $g_o\sim g_s^{1/2}\gg g_s$, the closed string effects are subdominant in the photon scattering. In particular, each sector contributes to the $F^4$ operators as\footnote{Here we assumed that the scale of compactification and volume of the cycles on which the brane wrapped are $\mathcal{O}(M_s)$. The same hierarchy is expected to hold as long as there is no unusual hierarchy between them.}
\begin{align}
[\alpha_{1,2}]_{\rm open}\sim \frac{M_{\rm Pl}^2}{g_sM_s^2}\,,
\quad
[\alpha_{1,2}]_{\rm closed}\sim \frac{M_{\rm Pl}^2}{M_s^2}\,,
\end{align}
and then unitarity implies
\begin{align}
\alpha_{1,2}\simeq[\alpha_{1,2}]_{\rm open}>0\,.
\end{align}
As an example, the positivity of $\alpha_1$ can explicitly be seen in the photon scattering of type-I superstring, where infinitely many higher spin open string states contribute to the effective coupling $\alpha_{1,2}$ (see also~footnote~\ref{footnote:positivity} and the Supplemental Material). Also recall that the graviton has to be accompanied by an infinite tower of higher spin particles, i.e., the Regge states, with the mass scale $m\sim M_{\rm Pl}/\alpha_3^{1/2}$ if $\alpha_3$ is nonzero. Indeed, in the bosonic string we have,
\begin{align}
\text{bosonic string: } \alpha_3\sim\frac{M_{\rm Pl}^2}{M_s^2}\,.
\end{align}
Note that $\alpha_3$ is prohibited in $\mathcal{N}\geq1$ supersymmetric (SUSY) theories because it generates the helicity amplitudes $\mathcal{M}(1^+,2^+,3^{+2})$ and $\mathcal{M}(1^-,2^-,3^{-2})$  incompatible with the SUSY Ward-Takahashi identity (see, e.g.,~\cite{Elvang:2013cua}):
\begin{align}
\text{SUSY: } \alpha_3=0\,.
\end{align}
Therefore, both in SUSY and non-SUSY cases, $\alpha_3$ is suppressed compared with the open string contributions to $\alpha_1$. Clearly, we have
\begin{align}
\alpha_1+\frac{1}{2}\alpha_3\simeq [\alpha_1]_{\rm open}>0\,.
\end{align}
More generally, the mass scale of the Regge states associated to the graviton is specified by the value of $\alpha_3$ (if nonzero) and their contribution to $\alpha_{1,2}$ is of the same order. If the photon is accompanied by another set of Regge states and these effects are dominant in the photon scattering, unitarity implies the inequality~\eqref{bound_on_alpha} and thus the mild form of the WGC is satisfied.

\section{Conclusion}

In this Letter, based on unitarity and causality, we demonstrated that heavy extremal BHs have the charge-to-mass ratio bigger than unity $z>1$ under some assumptions. The coverage of our argument is summarized by the flow chart Fig.~\ref{Fig:flow}. This provides an existence proof of the mild form of WGC in a wide class of theories, including generic stringy setups with dilaton or moduli stabilized below the string scale.

As a concluding remark, we present several promising future directions. First, while our proof already has wide applicabilities, it would be desirable to relax further the assumptions in the present work. A nontrivial example for the UV completion not covered by our argument is the heterotic superstring (with the stabilization scale $\gtrsim M_s$): Since both the photon and the graviton are from the closed string, we cannot directly apply our unitarity argument for $\alpha_1>0$ and $\alpha_2>0$. Nevertheless, from the explicit calculation of scattering amplitudes~\cite{Gross:1985rr}, we know that $\alpha_1$ and $\alpha_2$ are positive. Also, $\alpha_3=0$ because of SUSY. Hence, the inequality~\eqref{bound_on_alpha} is satisfied. Here we would like to remark that this observation is applicable as long as the tree-level scattering accommodates the same structure. For example, it is applicable to the heterotic superstring without spacetime SUSY~\cite{AlvarezGaume:1986jb}, where spacetime SUSY is broken by an unconventional GSO projection, but
the tree-level vertices of the bosonic sector are the same as the ordinary $E_8\times E_8$ heterotic superstring. 
It would be interesting to find out how the UV 
consistencies of string theory lead to the right sign.
We expect that besides unitarity and causality, some additional ingredients such as modular invariance may be behind  the validity of the WGC. We hope to revisit this issue elsewhere.

Another important direction is to extend our argument to other swampland conjectures. For example, it was conjectured in~\cite{Ooguri:2016pdq} that any non-supersymmetric AdS vacuum must be unstable (See \cite{Ibanez:2017kvh,Hamada:2017yji,Gonzalo:2018tpb,Gonzalo:2018dxi} for the application to the particle physics). Since the near-horizon geometry of an extremal BH is AdS, this conjecture is well motivated by our result showing that a decay process of extremal BHs is kinematically allowed. Further studies in this direction will be encouraged.
We believe that our work provides a foundation for such future studies in the swampland program and for deepening our understanding of the quantum gravity landscape.

\bigskip
\bigskip
\section*{Acknowledgments}

We would like to thank Panagiotis Betzios, Clifford Cheung, Elias Kiritsis, Hirosi Ooguri, Grant Remmen, and Cumrun Vafa for useful discussion. 
We also thank the Simons Center for Geometry and Physics Summer Workshop 2018, during which this work started.
Y.~H.  is supported in part by the Advanced ERC grant SM-grav, No 669288 and Grant-in-Aid for JSPS Fellows No.16J06151.
Y.~H. thanks the hospitality of the AstroParticule et Cosmologie (APC).
T.~N. is supported in part by JSPS KAKENHI Grant Numbers JP17H02894 and JP18K13539, and
MEXT KAKENHI Grant Number JP18H04352. 
G.~S. is supported in part by the DOE grant DE-SC0017647 and the Kellett Award of the University of Wisconsin.


\pagebreak
\widetext
\begin{center}
\textbf{\large Supplemental Material for ``Weak Gravity Conjecture from Unitarity and Causality"}
\end{center}
\setcounter{equation}{0}
\setcounter{figure}{0}
\setcounter{table}{0}
\setcounter{page}{1}
\makeatletter
\renewcommand{\theequation}{S\arabic{equation}}
\renewcommand{\thefigure}{S\arabic{figure}}

\medskip
This Supplemental Material includes derivation of the unitarity bounds~\eqref{unitarity_alpha} on the Wilson coefficients (Sec.\,I), extension of our argument to general spacetime dimension (Sec.\,II), implications from unitarity and causality on BH entropy corrections (Sec.\,III) and comments on Ref.~\cite{Cheung:2018cwt} (Sec.\,IV).

\section{I. Derivation of the unitarity bounds}

In this section we derive the positivity bounds,
\begin{align}
\label{unitarity_bound}
\alpha_1>0\,,\quad\alpha_2>0\,,
\end{align}
from unitarity assuming that Regge type massive states which UV complete gravity give subdominant contributions to the four-photon amplitudes compared to other massive intermediates states.

\subsection{A basis for forward scattering amplitudes}

To derive the unitarity bound, we first introduce a basis of forward scattering amplitudes consistent with factorization and are well-bounded at high energy.
Our starting point is the following spinning polynomial basis introduced in~\cite{Arkani-Hamed:2017jhn}:
\begin{align}
\nonumber
P_{\mathbf{s}}^{1234}(x)&=\frac{1}{(\mathbf{s}!)^2}\sum_a\frac{(\mathbf{s}+h_4-h_3)!(\mathbf{s}+h_3-h_4)!(\mathbf{s}+h_1-h_2)!(\mathbf{s}+h_2-h_1)!}{a!(\mathbf{s}+h_4-h_3-a)!(\mathbf{s}+h_2-h_1-a)!(a+h_1+h_3-h_2-h_4)!}
\\
\label{spb}
&\qquad\qquad\quad
\times\left(\frac{x-1}{2}\right)^{a+\frac{h_1+h_3-h_2-h_4}{2}}
\left(\frac{x+1}{2}\right)^{\mathbf{s}-a-\frac{h_1+h_3-h_2-h_4}{2}}\,,
\end{align}
where the summation over $a$ is from $\max\{0,-(h_1+h_3)+(h_2+h_4)\}$ to $\min\{\mathbf{s}+h_2-h_1,\mathbf{s}+h_4-h_3\}$.
The polynomial~\eqref{spb} is arranged such that the function,
\begin{align}
\label{partial_s}
\frac{g_{h_1h_2n}g_{\bar{h}_3\bar{h}_4n}}{m_n^2-s}P_{\mathbf{s}_n}^{1234}\left(1+\frac{2t}{m_n^2}\right)\,,
\end{align}
reproduces the $s$-channel factorization of four-point scattering amplitudes of massless particles with helicity $h_i$ (in the all-incoming notation). Here the intermediate massive state (labeled by $n$) carries the mass $m_n$ and the spin $\mathbf{s}_n$, and $g_{h_1h_2n}$ stands for the cubic coupling of massless particles with the helicity $h_1$ and $h_2$, and the massive state $n$. We also introduced $\bar{h}_i=-h_i$. In the forward limit, it simply reduces to
\begin{align}
\label{forward_s}
\frac{g_{h_1h_2n}g_{\bar{h}_3\bar{h}_4n}}{m_n^2-s}P_{\mathbf{s}_n}^{1234}\left(1\right)
\end{align}
with an explicit form,
\begin{align}
\label{P_simplified1}
P_{\mathbf{s}}^{1234}\left(1\right)=\frac{(\mathbf{s}+h_1-h_2)!(\mathbf{s}+h_2-h_1)!}{(\mathbf{s}!)^2}
\end{align}
for $h_1-h_2=h_4-h_3$, and 
\begin{align}
P_{\mathbf{s}}^{1234}(1)=0
\end{align}
for $h_1-h_2\neq h_4-h_3$. Here one might worry that Eq.~\eqref{P_simplified1} is singular for $\mathbf{s}<|h_1-h_2|$. However, the cubic coupling $g_{h_1h_2n}$ for $\mathbf{s}_n<|h_1-h_2|$ is prohibited by locality~\cite{Arkani-Hamed:2017jhn}, so that Eq.~\eqref{forward_s} is always regular off shell.

We then introduce a basis for forward scattering amplitudes of massless particles. For the moment, let us assume that there is no massless particle exchange and neglect gravitational effects in particular. Later we will discuss under which conditions this assumption may be justified when working in gravitational systems. Under this assumption, scattering amplitudes are finite in the forward limit $t\to0$ and thus they depend only on $s$. Since their non-analytic structure can completely be determined by the factorization property, we may write
\begin{align}
\mathcal{M}^{1234}(s)
=\sum_n
\bigg[
\frac{g_{h_1h_2n}g_{\bar{h}_3\bar{h}_4n}}{m_n^2-s}P_{\mathbf{s}_n}^{1234}(1)
+\frac{g_{h_1h_4n}g_{\bar{h}_3\bar{h}_2n}}{m_n^2+s}P_{\mathbf{s}_n}^{1432}(1)
\bigg]
+\text{terms analytic in $s$}\,,
\end{align}
where the first and the second terms are responsible for the non-analyticity in the $s$- and $u$-channels. Also, the summation over the intermediate states $n$ may be either discrete or continuous. For example, the loop effects generate a branch cut, which can be understood as an infinite sum of single poles. Here let us assume that the forward amplitude is suppressed as $<s^2$ in the UV limit. It is the case, e.g., when the amplitude satisfies the Froissart bound.
We may then specify the analytic part as\footnote{We assume that the summation over $n$ is absolutely convergent, so that the summation and the limit $s\to\infty$ are commutative.}
\begin{align}
\label{basis_forward}
\mathcal{M}^{1234}(s)
=\sum_n
\bigg[
\frac{g_{h_1h_2n}g_{\bar{h}_3\bar{h}_4n}}{m_n^2-s}P_{\mathbf{s}_n}^{1234}(1)
+\frac{g_{h_1h_4n}g_{\bar{h}_3\bar{h}_2n}}{m_n^2+s}P_{\mathbf{s}_n}^{1432}(1)
+a_n+b_ns
\bigg]
\,,
\end{align}
where $a_n$ and $b_n$ are numerical constants.

\subsection{Positivity bounds}

We then apply this basis to derive the bound~\eqref{unitarity_bound}. For this purpose, let us first recall that the forward helicity amplitudes generated by the effective interactions~\eqref{EFT} are
\begin{align}
\mathcal{M}^{\pm\pm\pm\pm}(s)
=\frac{4}{M_{\rm Pl}^4}(\alpha_1-\alpha_2)s^2
\,,
\quad
\mathcal{M}^{\pm\pm\mp\mp}(s)
=\mathcal{M}^{\pm\mp\mp\pm}(s)
=\frac{2}{M_{\rm Pl}^4}(\alpha_1+\alpha_2)s^2
\,,
\quad
\text{(others)}=0\,,
\end{align}
where we neglected graviton effects based on the aforementioned assumption. Also the label $\pm$ indicates that the external particle has a helicity $\pm1$. We may then obtain the relation between the Wilson coefficients $\alpha_i$ and the UV data $(g_n,m_n)$ by expanding Eq.~\eqref{basis_forward} in $s$ and reading off the coefficient of $s^2$. Assuming parity invariance,  i.e., $g_{++n}=g_{--n}$ for party even intermediate states and $g_{++n}=-g_{--n}$ for parity odd ones, we find\footnote{Note that the $\mathcal{O}(s^0)$ and $\mathcal{O}(s^1)$ terms can be set to zero by appropriately choosing $a_n$ and $b_n$ in Eq.~\eqref{basis_forward}.}
\begin{align}
\frac{4\alpha_1}{M_{\rm Pl}^4}=
\sum_{n:\text{ even}}
\frac{2g_{++n}^2}{m_n^6}
+\sum_n\frac{(\mathbf{s}_n\!+\!1)(\mathbf{s}_n\!+\!2)}{\mathbf{s}_n(\mathbf{s}_n\!-\!1)}\frac{g_{+-n}^2}{m_n^6}\,,
\qquad
\frac{4\alpha_2}{M_{\rm Pl}^4}=
\sum_{n:\text{ odd}}
\frac{2g_{++n}^2}{m_n^6}
+\sum_n\frac{(\mathbf{s}_n\!+\!1)(\mathbf{s}_n\!+\!2)}{\mathbf{s}_n(\mathbf{s}_n\!-\!1)}\frac{g_{+-n}^2}{m_n^6}\,,
\end{align}
where $\sum_{n:\text{even}}$ stands for the summation over parity even intermediate states and similarly for $\sum_{n:\text{odd}}$. Note that the spin statistics and locality imply that $g_{++n}=0$ for odd spin $\mathbf{s}_n$ and $g_{+-n}=0$ for $\mathbf{s}_n=0,1$~\cite{Arkani-Hamed:2017jhn}. We therefore conclude that $\alpha_1>0$ ($\alpha_2>$0) follows from unitarity if the photon is coupled to a parity even (odd) scalar or a spin $s\geq2$ state with an arbitrary parity.

\subsection{On graviton exchange and Regge states}

In the rest of this section, we discuss how graviton affects the argument and under which conditions the bounds~\eqref{unitarity_bound} may be justified. A crucial obstruction for deriving a rigorous unitarity bound in gravitational systems is that the $t$-channel graviton exchange dominates in the forward limit $t\to 0$ and quadratically diverges in the high energy limit:
\begin{align}
\mathcal{M}(s,t)\sim -\frac{1}{M_{\rm Pl}^2}\frac{s^2}{t}
\,.
\end{align}
Then, the ansatz~\eqref{basis_forward} is not applicable anymore\footnote{Nevertheless, it is interesting to notice that in the setup~\eqref{EFT}, the average of all the helicity amplitudes goes in the forward limit as
\begin{align}
\mathcal{M}(s,t)\simeq\frac{-1}{4M_{\rm Pl}^2}\left(\frac{s^2}{t}+s\right)+\frac{2\alpha_1-\alpha_3}{2M_{\rm Pl}^2}s^2\,.
\end{align}
The bound~\eqref{bound_on_alpha} then directly follows from the positivity of the $\mathcal{O}(s^2)$ coefficient {\it if} we can subtract the divergent contribution from the graviton exchange in a controllable way. While our paper provides a solid criterion to satisfy the bound~\eqref{bound_on_alpha}, it would be interesting to explore some technique to read off the sign of non-singular part of the $\mathcal{O}(s^2)$ term in more general contexts.}. Presumably, the only way to make mild the UV behavior of forward scattering is for the graviton to be accompanied by higher spin states (Regge states) to have amplitudes of the form,
\begin{align}
\label{Regge}
\mathcal{M}(s,t)\sim -\frac{1}{M_{\rm Pl}^2}\frac{s^{2+\gamma t+\mathcal{O}(t^2)}}{t}\,,
\end{align}
in the Regge limit $s\to\infty$ ($t<0$ : fixed). The amplitude is then bounded as $<s^2$ for small negative $t$ as long as $\gamma>0$. It is also instructive to expand Eq.~\eqref{Regge} as
\begin{align}
\mathcal{M}(s,t)\sim -\frac{1}{M_{\rm Pl}^2}\frac{s^2}{t}-\frac{\gamma}{M_{\rm Pl}^2}s^2\log s+\mathcal{O}(t)\,,
\end{align}
which implies that the $\mathcal{O}(t^0)$ term behaves as $\sim s^2\log s$ in the high energy limit. It is polynomially bounded $<s^3$, but less bounded than the non-gravitational case. Hence, $\mathcal{O}(s^2)$ terms may appear as an analytic component of Eq.\eqref{basis_forward}. Since these new analytic $\mathcal{O}(s^2)$ terms may have a negative coefficient, we can not derive the positivity bound on the $\mathcal{O}(s^2)$ term anymore.

Does it mean that we cannot say anything about the sign of the Wilson coefficients $\alpha_i$ in gravitational systems? Physically speaking, it is not true: We should be able to get bounds at least when the gravitational effects are small enough. For example in string theory, the coefficient $\gamma$ is determined by the string scale $M_s$ as $\gamma\sim M_s^{-2}$ and the $\mathcal{O}(t^0)$ component of the scattering amplitude is schematically of the form,
\begin{align}
\frac{s^2}{M_{\rm Pl}^2M_s^2}\sum_{n=0}^\infty c_n\left(\frac{s}{M_s^2}\right)^n\,,
\end{align}
with $\mathcal{O}(1)$ coefficients $c_n$. Then, the Regge states (associated to the graviton and thus are closed strings) contribute to $\alpha_1$ and $\alpha_2$ as
\begin{align}
\label{alpha_Regge}
[\alpha_{1,2}]_{\rm Regge}\sim \frac{M_{\rm Pl}^2}{M_s^2}\,.
\end{align}
The bound~\eqref{unitarity_bound} is therefore applicable if there exists other intermediate states generating four-photon effective interactions bigger than Eq.~\eqref{alpha_Regge}. This is the case, e.g., when the photon is coupled to a light neutral boson with the mass $m\ll M_s$. Another typical example is when the photon comes from an open string. Since the intermediate open string states generate effective interactions,
\begin{align}
[\alpha_{1,2}]_{\rm open}\sim\frac{M_{\rm Pl}^2}{g_sM_s^2}\,,
\end{align}
this effect is parametrically bigger than~\eqref{alpha_Regge} due to a factor $g_s^{-1}\gg1$. Note that the infinitely many higher spin open string states contribute to the effective coupling $\alpha_{1,2}$ (recall that intermediate states with an arbitrary spin may contribute). More generally, we expect that the bound~\eqref{unitarity_bound} may be justified if the contribution,
\begin{align}
\label{Regge_alpha}
[\alpha_{1,2}]_{\rm Regge}\sim \frac{M_{\rm Pl}^2}{\Lambda_{\rm Regge}^2}\,,
\end{align}
from the Regge states which UV complete gravity is subdominant compared to other effects, where we introduced the typical mass scale of the Regge states as $\Lambda_{\rm Regge}$.

\section{II. Extension to $D\geq5$}

In this section we extend our $4D$ argument to general spacetime dimension $D\geq5$. The WGC bound in general dimension $D$ is stated as
\begin{align}
\label{WGC_D}
z=\sqrt{\frac{D-2}{D-3}}\frac{|q|}{\kappa m}\geq1\,.
\end{align}
Here $\kappa$ is the gravitational coupling, whose normalization is given shortly. Below we show that the heavy extremal BHs may satisfy the WGC bound~\eqref{WGC_D} under similar assumptions of the $4D$ argument in the main text based on unitarity and causality.

\subsection{Effective action and unitarity}

In general dimension $D$, the general effective action for the photon and the graviton is given up to four derivatives by \begin{align}
S&=\int d^Dx\sqrt{-g}
\bigg[
\frac{R}{2\kappa^2}-\frac{1}{4}F_{MN}F^{MN}
+\kappa^2\Big(a_1(F_{MN}F^{MN})^2
+a_2F_{MN}F^{NR}F_{RS}F^{SM}\Big)
\nonumber
\\
\label{EFT_D}
&\qquad\qquad\qquad\quad
+a_3F_{MN}F_{RS}W^{MNRS}
+\frac{a_4}{\kappa^2}W_{MNRS}W^{MNRS}
\bigg]\,,
\end{align}
where $\kappa$ is the gravitational coupling\footnote{
For notational simplicity, we use the gravitational coupling $\kappa$ rather than the Planck mass $M_{\rm Pl}$. Also for the same reason we use the Wilson coefficients $a_i$ of  mass dimension $-2$.}. In contrast to the $D=4$ case, we have four independent higher derivative operators $a_i$. One of them is to account for the Gauss-Bonnet term, which is topological and thus does not affect the equations of motion in $D=4$.
Also, in $D=4$, the $a_2$ operator in Eq.~\eqref{EFT_D} is equivalent to a linear combination of the $\alpha_1$ and $\alpha_2$ operators in Eq.~\eqref{EFT}. More explicitly, if we focus on scattering processes on a $4D$ subspacetime (labeled by Greek indices),  i.e., if all the polarization and momentum vectors are constrained on it, the dynamics is captured by the effective Lagrangian,
\begin{align}
\mathcal{L}&=\frac{1}{2\kappa^2}R-\frac{1}{4}F_{\mu\nu}F^{\mu\nu}
+\kappa^2\left[
\left(a_1+\frac{a_2}{2}\right)(F_{\mu\nu}F^{\mu\nu})^2
+\frac{a_2}{4}(F_{\mu\nu}\widetilde{F}^{\mu\nu})^2\right]
+a_3F_{\mu\nu}F_{\rho\sigma}W^{\mu\nu\rho\sigma}
+\frac{a_4}{\kappa^2}W_{\mu\nu\rho\sigma}W^{\mu\nu\rho\sigma}\,,
\end{align}
where we used the following identity applicable in $D=4$:
\begin{align}
F_{\mu\nu}F^{\nu\rho}F_{\rho\sigma}F^{\sigma\mu}
=\frac{1}{2}(F_{\mu\nu}F^{\mu\nu})^2+\frac{1}{4}(F_{\mu\nu}\widetilde{F}^{\mu\nu})^2\,.
\end{align}
Since the Weyl tensor does not couple to the matter sector in the Einstein equation, the four-point contact vertices of photon are from the $(F_{\mu\nu}F^{\mu\nu})^2$ and $(F_{\mu\nu}\widetilde{F}^{\mu\nu})^2$ terms only. Note that if we write the effective action in terms of the Riemann tensor instead of the Weyl tensor, its Ricci tensor component may generate four-point photon vertices through the Einstein equation.
By analyzing the unitarity of this scattering process on the $4D$ subspacetime, we may derive essentially the same unitarity constraints as in Table~\ref{table} on $a_i$. We now have
\begin{align}
\label{unitarity_D}
a_1+\frac{a_2}{2}>0\,,
\quad
a_2>0
\end{align}
for the tree-level effect (a) from light neutral bosons and the loop effect (b-1) from light charged particles with $z\gg1$ (more generally, those coupled to the photon by some interactions stronger than gravity). {Also, if the photon and the graviton are accompanied by different sets of Regge states and those for graviton are subdominant in the photon scattering, the UV effect (c) enjoys~\eqref{unitarity_D}. For the same reason as the $D=4$ case, we focus on the tree-level effects (a) and (c) in the following.

\subsection{WGC from unitarity and causality} 

We then discuss implications of unitarity on the WGC. In general spacetime dimension $D$, the charge-to-mass ratio of heavy extremal BHs is given by~\cite{Kats:2006xp}
\begin{align}
\label{z_D}
z=\sqrt{\frac{D-2}{D-3}}\frac{|Q|}{\kappa M}
=1+\left(
\frac{(D-2)(D-3)\Omega_{D-2}^2}{\kappa^2 Q^2}
\right)^{1/(D-3)}\!\!\!\mathcal{F}(a_i)\,,
\end{align}
where $\Omega_{D-2}$ is the area of $S_{D-2}$ with a unit radius:
\begin{align}
\label{Omega}
\Omega_{D-2}=\frac{2\pi^{(D-1)/2}}{\Gamma(\frac{D-1}{2})}\,.
\end{align}
$\mathcal{F}(\alpha_i)$ is a function of the Wilson coefficients $a_i$ given by
\begin{align}
\label{bound_D}
\mathcal{F}(a_i)
&=\frac{2(D-2)(D-3)^2}{(3D-7)}(2a_1+a_2)
+\frac{2(D-3)^2(D^2-8D+13)}{(D-1)(3D-7)}
a_3
+\frac{4(D-3)(D^3-9D^2+28D-29)}{(D-1)(3D-7)} a_4\,,
\end{align}
where in particular the numerical coefficients of $2a_1+a_2$ and $a_4$ in Eq.~\eqref{bound_D}  are positive for any $D\geq4$.
Heavy extremal BHs then have the charge-to-mass ratio $z\geq1$ and the mild form of WGC is satisfied if $\mathcal{F}(a_i)\geq0$. 
\subsubsection{Causality constraints}

Similar to the $D=4$ case, the operators $a_3$ and $a_4$ lead to causality violation at the energy scale where these corrections become relevant. Again, an infinite tower of massive higher spin particles, i.e., the Regge states associated to the graviton, is required at this scale to UV complete the EFT at  tree-level without causality violation~\cite{Camanho:2014apa} (see also~\cite{Li:2017lmh,Afkhami-Jeddi:2016ntf} for a holographic derivation based on the conformal bootstrap approach). Then, all the tree-level contributions to $a_3$ and $a_4$ are classified into the UV effect (c) in Table~\ref{table}.

\paragraph{Case (1): light neutral bosons}

The causality constraints imply that  if the tree-level effect (a) of light neutral bosons is the dominant effect, $a_3$ and $a_4$ are negligible. Together with unitarity, we find
\begin{align}
\label{F_D>4}
\mathcal{F}(a_i)
&\simeq\frac{2(D-2)(D-3)^2}{(3D-7)}(2a_1+a_2)>0\,.
\end{align}
The mild form of the WGC is then satisfied by heavy extremal BHs even if there are no charged particles with $z\geq1$, as long as a light neutral particle is coupled to the photon and the effect (a) dominates over the others.

\paragraph{Case (2): open string type UV completion}

We can also see that the open string type UV completion enjoys the inequality $\mathcal{F}(a_i)>0$. First, causality implies that $a_{3,4}$ (if nonzero) specifies the mass scale $\Lambda_{\rm Regge}$ of the Regge states associated to the graviton as
\begin{align}
a_3\sim \frac{1}{\Lambda_{\rm Regge}^2}
\,,
\quad
a_4 \sim \frac{1}{\Lambda_{\rm Regge}^2}\,.
\end{align}
These Regge states for the graviton contribute to $a_{1,2}$ as
\begin{align}
[a_{1,2}]_{\rm Regge}\sim \frac{1}{\Lambda_{\rm Regge}^2}
\,.
\end{align}
Hence, if the photon is accompanied by another set of Regge states and they dominate over the graviton Regge states in the photon scattering, we have a hierarchy,
\begin{align}
|a_{1,2}|\gg |a_{3,4}|\,,
\end{align}
and the bounds~\eqref{unitarity_D}. As a result, the inequality $\mathcal{F}(a_i)>0$ and thus the mild form of the WGC are satisfied.

Note that $a_3$ and $a_4$ are constrained by SUSY also. As in the 4D case, the $a_3$ operator $F_{MN}F_{RS}W^{MNRS}$ is prohibited by $\mathcal{N}=1$ SUSY. On the other hand, the $a_4$ operator $W^2$, which has the same coefficient as the Gauss-Bonnet term, is incompatible with maximum SUSY ($\mathcal{N}=8$ in terms of $D=4$), even though it is compatible with half maximal SUSY.

\section{III. Positivity of BH entropy correction}

In this section we calculate the higher derivative correction to the BH entropy and show that it is positive for any charged BH in the two classes of theories studied in our paper, as a consequence of unitarity and causality.
 
\subsection{Entropy formula}

First, for spherically symmetric BHs with the metric,
\begin{align}
ds^2=-f(r)dt^2+\frac{dr^2}{g(r)}+r^2d\Omega_{D-2}^2\,,
\end{align}
the Wald entropy formula reads~\cite{Wald:1993nt}\footnote{We use the same symbol $S$ for the entropy as the action, but it will be obvious from the context which we refer to.}
\begin{align}
S=-2\pi\Omega_{D-2} r_H^{D-2}\left[\frac{\partial \mathcal{L}}{\partial R_{MNRS}}\epsilon_{MN}\epsilon_{RS}\right]_{\rm horizon}\,,
\end{align}
where $\Omega_{D-2}$ is the area of $S_{D-2}$ with a unit radius. See Eq.~\eqref{Omega}. $r_H$ is the radius of the event horizon (the outer horizon for charged BHs) and $[...]_{\rm horizon}$ denotes a value evaluated on it. $\epsilon_{MN}$ is the binormal on the event horizon normalized as $\epsilon_{MN}\epsilon^{MN}=-2$. In our setup~\eqref{EFT_D},
the BH entropy can then be calculated as
\begin{align}
S=\frac{2\pi\Omega_{D-2}r_H^{D-2}}{\kappa^2}+\Delta S_{\rm Wald}\,,
\end{align}
where the first term reproduces the area low of the Einstein-Maxwell theory and the second term $\Delta S_{\rm Wald}$ is the higher derivative correction to the area low given by
\begin{align}
\Delta S_{\rm Wald}&=
2\pi\Omega_{D-2}r_H^{D-2}\bigg[-a_3\kappa^2\left(F^{MN}F^{RS}\epsilon_{MN}\epsilon_{RS}-\frac{4}{D-2}F^{ML}F^R{}_Lg^{NS}\epsilon_{MN}\epsilon_{RS}-\frac{4}{(D-1)(D-2)}F_{AB}F^{AB}\right)
\nonumber
\\
&\qquad\qquad\qquad\qquad
-2a_4W^{MNRS}\epsilon_{MN}\epsilon_{RS}
\bigg]_{\rm horizon}\,.
\end{align}
Note that the BH solution is modified by higher derivative operators and so is the horizon radius $r_H$. The first area law term is therefore corrected by higher derivative corrections.
Therefore, the leading correction to the entropy formula is given by
\begin{align}\label{Eq:entropy_contributions}
S=S_{\rm EM}+\Delta S
\quad
{\rm with}
\quad
\Delta S=\Delta S_{\rm horizon}+\Delta S_{\rm Wald}\,,
\end{align}
where $\Delta S$ is the leading higher derivative correction to the BH entropy $S_{\rm EM}$ in the Einstein-Maxwell theory. $\Delta S_{\rm horizon}$ is the effect of the horizon shift.
After straightforward but somewhat tedious calculations, we find an explicit form of $\Delta S_{\rm Wald}$ as
\begin{align}
\frac{\Delta S_{\rm Wald}}{S_{\rm EM}}=\frac{1}{(3D-7)m^{\frac{2}{D-3}}\xi(1+\xi)^{\frac{D-1}{D-3}}}
&\bigg[
-\frac{4(1-\xi)(D-2)(D-3)}{D-1} (D-3)(3D-7) \xi a_3
\nonumber \\
\label{S_Wald}
&\quad
-\frac{8(D\!-\!2)}{D\!-\!1}(D\!-\!3)(3D\!-\!7)\big((D\!-\!4)-(2D\!-\!5)\xi\big) \xi a_4
\bigg]\,,
\end{align}
while $\Delta S_{\rm horizon}$ is given by
\begin{align}
\nonumber
\frac{\Delta S_{\rm horizon}}{S_{\rm EM}}
&=
\frac{1}{(3D\!-\!7)m^{\frac{2}{D\!-\!3}}\xi(1+\xi)^{\frac{D\!-\!1}{D\!-\!3}}}
\\
&\quad
\times\bigg[
2(D\!-\!2)^2(D\!-\!3)(1\!-\!\xi)^2(2a_1+a_2)
+\frac{2(1\!-\!\xi)(D\!-\!2)(D\!-\!3)}{D\!-\!1} 
\big(
(D^2\!-\!8D+13)
+2(D^2\!-\!7D+11)\xi
\big)a_3
\nonumber\\
\label{S_horizon}
&\qquad\quad
+\frac{4(D\!-\!2)}{D\!-\!1}
\bigg(
(D^3\!-\!9D^2+28D\!-\!29)(1\!-\!\xi)^2
+(3D\!-\!7)\xi\big((2D^2\!-\!13D+23)\!-\!3(D\!-\!3)^2\xi\big)
\bigg)a_4
\bigg]\,.
\end{align}
Here, instead of the BH mass $M$ and the charge $Q$,
we used the rescaled ones, $m$ and $q$, defined by
\begin{align}
m=\frac{\kappa^2M}{(D\!-\!2)\Omega_{D-2}}
\,,
\quad
q=\frac{\kappa Q}{\sqrt{(D\!-\!2)(D\!-\!3)}\Omega_{D-2}}\,,
\end{align}
such that the extremality condition is $|q|=m$ in the Einstein-Maxwell theory. We also introduced
\begin{align}
\xi=\sqrt{1-\frac{q^2}{m^2}}\,,
\end{align}
so that $\xi=1$ for Schwarzschild BHs and $\xi=0$ for extremal BHs in the Einstein-Maxwell theory.

\subsection{Positivity for generic charged BHs} 

By adding two contributions in \eqref{Eq:entropy_contributions}, the higher derivative correction to the BH entropy in the setup~\eqref{EFT_D} reads
\begin{align}
\label{DeltaS}
\frac{\Delta S}{S_{EM}}=\frac{1}{(3D-7)m^{\frac{2}{D-3}}\xi(1+\xi)^{\frac{D-1}{D-3}}}
\mathcal{G}(a_i,\xi)
\end{align}
with a function $\mathcal{G}(a_i,\xi)$ given by
\begin{align}
\mathcal{G}(a_i,\xi)&=2(D-2)^2(D-3)(1-\xi)^2(2a_1+a_2)
\nonumber
\\
&\quad
+\frac{2(1-\xi)(D-2)(D-3)}{D-1}\left[(D^2- 8 D +13) - 2(D-2)(2D-5) \xi\right]a_3
\nonumber
\\
\label{calG}
&\quad
+\frac{4(D-2)}{D-1}
\Big[
(D^3-9D^2+28D-29)(1-\xi)^2
+(D-1)(3D-7)\xi\big(1+(D-3)\xi\big)
\Big]a_4\,.
\end{align}
Note that Eq.~\eqref{DeltaS} is singular in the extremal limit $\xi=0$, but this divergence is not physical: $\xi\simeq0$ is simply out of validity of the approximation. We provide the entropy correction formula for $\xi=0$ later.

It is easy to see that the coefficients of $2a_1+a_2$ and $a_4$ in Eq.~\eqref{calG} are positive for any $0<\xi<1$ and $D\geq4$. Also recall that the two setups we considered accommodate the hierarchy,
\begin{align}
|a_1|,\,|a_2|&\gg|a_3|,\,|a_4|\qquad{\rm for}\quad D\geq5\,,
\\
|a_1|,\,|a_2|&\gg|a_3|\qquad\qquad\,{\rm for}\quad D=4,
\end{align}
and satisfy the positivity bound,
\begin{align}
2a_1+a_2>0\,,
\end{align}
as a consequence of unitarity and causality. Hence, in $D\geq5$, the function $\mathcal{G}(a_i,\xi)$ and therefore the entropy correction are positive for any charged BH in the two setups ($\xi=0$ is discussed separately later). Note that the entropy correction for the Schwarzschild BHs $\xi=1$ is determined only by the coefficient $a_4$ of the Gauss-Bonnet term and the entropy correction is positive if and only if the coefficient is positive $a_4>0$. To our knowledge, no rigorous unitarity proof of this bound is known so far\footnote{
In~\cite{Cheung:2016wjt} it was claimed that the positivity of the Gauss-Bonnet term follows from unitarity by a spectral decomposition argument similar to~\cite{Cheung:2014ega} (see also footnote~\ref{footnote:positivity}). However, the interaction considered there is restrictive to the tree-level exchange of heavy states with the same index structure as the Weyl tensor and the interaction is singular in the UV. Moreover, the Gauss-Bonnet term contains a cubic graviton interaction, which can be thought of as a fundamental vertex. It is not clear how such a cubic coupling arises from tree-level exchange of heavy particles.}}.

We also remark that in $D=4$, the Gauss-Bonnet term is topological and thus it does not affect the equations of motion. However, it is known to give an entropy correction proportional to the Euler number of the horizon (see, e.g.,~\cite{Jacobson:1993xs,Liko:2007vi,Sarkar:2010xp}). Since our argument is based on unitarity and causality of scattering amplitudes, we cannot constrain the topological contribution from $a_4$. However, if we assume that the Gauss-Bonnet term in $D=4$ comes from that in $D\geq5$, causality in the higher dimension requires that its contribution is negligible in the aforementioned two setups. Under this assumption, we can derive $\mathcal{G}(a_i,\xi)>0$ and thus $\Delta S>0$ in $D=4$ from unitarity and causality. Besides, it has been argued that the topological contribution to the BH entropy potentially leads to the second law violation~\cite{Jacobson:1993xs,Liko:2007vi,Sarkar:2010xp}. Even if the Gauss-Bonnet term has no higher dimensional origin, it will not be unreasonable to simply assume that $a_4$ is negligible to avoid the potential second law violation. In this case, $\Delta S>0$ in $D=4$ again follows from unitarity and causality in our two setups.

\subsection{Positivity for extremal BHs}

We then calculate the entropy correction for $\xi=0$ and demonstrate that the correction to the extremality condition is directly related to the entropy correction for BHs with $\xi=0$ saturating the extremal bound of the Einstein-Maxwell theory. As we mentioned, the leading order correction to the BH entropy can schematically be written as
\begin{align}
\Delta S=\Delta S_{\rm horizon}+\Delta S_{\rm Wald}\,,
\end{align}
where $\Delta S_{\rm Wald}$ is from the higher derivative correction to Wald's entropy formula and it is $\mathcal{O}(a_i)$ for general $\xi$. See Eq.~\eqref{S_Wald}. On the other hand, $\Delta S_{\rm horizon}$ is due to the horizon shift $\Delta r_H$ by the higher derivative correction. More explicitly, it follows from the area low in the Einstein-Maxwell theory as
\begin{align}
\frac{\Delta S_{\rm horizon}}{S_{EM}}=\frac{(r_H+\Delta r_H)^{D-2}}{r_H^{D-2}}-1\simeq (D\!-\!2)\frac{\Delta r_H}{r_H}\,,
\end{align}
where $r_H$ is the horizon radius in the Einstein-Maxwell theory. For a not too small $\xi$, $\Delta S_{\rm horizon}$ is $\mathcal{O}(a_i)$ and its concrete form is given in Eq.~\eqref{S_horizon}. For $\xi\simeq0$, however,  it is $\mathcal{O}(a_i^{1/2})$ as we calculate shortly, so that $|\Delta S_{\rm horizon}|\gg |\Delta S_{\rm int}|$ as long as higher derivative corrections are small (which is true for sufficiently heavy BHs as we mentioned in the main text). In this regime, the entropy correction is positive $\Delta S>0$ if the horizon shift $\Delta r_H$ is positive.

An immediate conclusion here is that the entropy correction for heavy BHs with $\xi=0$ is positive if the higher derivative corrections resolve the degeneracy of the two horizons without introducing a naked singularity: In this case, there appear two horizons at $r=r_H\pm\Delta r_H$, where the plus (minus) sign is for the outer (inner) horizon. The horizon shift for the outer horizon is then always positive and so is the entropy correction. Note that this condition is nothing but the requirement that the charge-to-mass ratio $z_{\rm ext}$ of heavy extremal BHs is shifted as $z_{\rm ext}>1$ (otherwise the modified BH solution has a naked singularity) and thus heavy extremal BHs play the role of the state with $z>1$ required by the mild form of WGC.

More explicitly, the entropy correction for $\xi=0$ is calculated as follows:
First, suppose that the BH solution takes the form,
\begin{align}
ds^2=-f(r)dt^2+\frac{dr^2}{g(r)}+r^2d\Omega_{D-2}^2\,,
\end{align}
after including higher derivative corrections. The location of the horizons is then determined by $f(r)=g(r)=0$. When higher derivative corrections are small, it is convenient to decompose $g(r)$ as
\begin{align}
g(r)=g_{EM}(r)+\Delta g(r)\,,
\end{align}
where $g_{EM}(r)$ is $g(r)$ in the Einstein-Maxwell theory:
\begin{align}
g_{EM}(r)=1-\frac{2m}{r^{D-3}}+\frac{q^2}{r^{2(D-3)}}\,,
\end{align}
where an explicit form of $\Delta g(r)$ is given by
\begin{align}
\Delta g(r)&=4{(d-3)^2\over d-1}{1\over r^2}
\bigg[
2{(D-1)(D-4)\over D-3}a_4{m^2\over r^{2(D-3)}}
-2\Big((D-3)a_3+2(2D-5)a_4\Big){q^2\over r^{2(D-3)}}
\nonumber
\\
&\quad
+\Big((3D-7)a_3+2{4D^2-21D+29\over D-3}a_4\Big){m\,q^2\over r^{3(D-3)}}
\nonumber
\\
&\quad
-{1\over 3D-7}\bigg\{
(D-1)(D-2)(2a_1+a_2)
+2(D-2)(2D-5)a_3
+2{(D-2)(2D-5)^2 \over D-3}a_4
\bigg\}{q^4\over r^{4(D-3)}}
\bigg].
\end{align}
We can then evaluate the horizon shift $\Delta r_H$ for generic $\xi$ by solving
\begin{align}
0&=g(r_H+\Delta r_H)
\simeq \Delta g(r_H)+\Delta r_H\,g'_{EM}(r_H)\,,
\end{align}
where we used $g_{EM}(r_H)=0$. This is how~\cite{Cheung:2018cwt} evaluated $\Delta S_{\rm horizon}$ for generic $\xi$. However, $g'_{EM}(r_H)$ vanishes for the extremal limit $\xi\to0$ because of the horizon degeneracy. We need then take into account the next order in the $\Delta r_H$ expansion:
\begin{align}
0=\Delta g(r_H)+\frac{1}{2}\Delta r_H^2\,g''_{EM}(r_H)
\quad
{\rm for}
\quad
\xi=0\,.
\end{align}
We then find
\begin{align}
\label{Delta_rH_extermal}
\frac{\Delta r_H^2}{r_H^2}=-\frac{2\Delta g(r_H)}{r_H^2g''_{EM}(r_H)}=\frac{4\mathcal{F}(a_i)}{(D-3)^2m^{\frac{2}{D-3}}}\,,
\end{align}
where $\mathcal{F}(a_i)$ is the same function as Eq.~\eqref{bound_D}. If $\mathcal{F}(a_i)>0$, the BH with $\xi=0$ has no naked singularity, but rather it has two horizons after including higher derivative corrections. The positive (negative) solution for Eq.~\eqref{Delta_rH_extermal} is for the outer (inner) horizon. Note that this is essentially the same statement that there exist BHs of the charge-to-mass ratio $z>1$ without naked singularity. Also, we see that the horizon shift is $\Delta r_H\propto \mathcal{F}(a_i)^{1/2}=\mathcal{O}(a_i^{1/2})$. The entropy correction is then
\begin{align}
\frac{\Delta S}{S_{EM}}\simeq\frac{\Delta S_{\rm horizon}}{S_{EM}}=\frac{2(D-2)}{(D-3)m^{\frac{1}{D-3}}}\sqrt{\mathcal{F}(a_i)}\,.
\end{align}
We have now explicitly shown that the entropy correction for $\xi=0$ is positive when $\mathcal{F}>0$ and thus heavy extremal BHs satisfy the WGC bound (which is the case, e.g., for the two setups we considered in this paper).

\section{IV. Comments on \cite{Cheung:2018cwt}}

Here we point out a loophole in the entropy argument given in~\cite{Cheung:2018cwt}.
As an illustrative example, let us consider the following Lagrangian of a massive spin 2 field, $h_{\mu\nu}$, coupled with $F^2$ term (see, e.g., the review~\cite{deRham:2014zqa} for the kinematics of a massive spin 2 field):
\begin{align}
\label{counterexample}
\mathcal{L}&=\mathcal{L}_{\rm EM}+\Delta\mathcal{L}\,,
\quad{\rm with}\quad
\Delta\mathcal{L}=-\frac{1}{4}h^{\mu\nu}\mathcal{E}_{\mu\nu}^{\alpha\beta}h_{\alpha\beta}-\frac{m^2}{8}(h_{\mu\nu}^2-h^2)
+{1\over M}h F_{\rho\sigma} F^{\rho\sigma}\, ,
\end{align}
where $\mathcal{L}_{\rm EM}$ is the Einstein-Maxwell Lagrangian and we introduced $h:=h^\mu_\mu$. The kinetic operator $\mathcal{E}_{\mu\nu}^{\alpha\beta}$ is\footnote{For simplicity, we neglected metric fluctuations around the Minkowski background. However, our conclusion up to the four derivative operators, which is relevant to the WGC argument, does not change even if we take into account their effects.}
\begin{align}
\mathcal{E}_{\mu\nu}^{\alpha\beta}h_{\alpha\beta}
=-\frac{1}{2}
\bigg[
\Box h_{\mu\nu}-\partial_\mu\partial_\alpha h^\alpha_\nu
-\partial_\nu\partial_\alpha h^\alpha_\mu
+\partial_\mu\partial_\nu h
-\eta_{\mu\nu}(\Box h-\partial_\alpha\partial_\beta h^{\alpha\beta})
\bigg]\,.
\end{align}
Since the trace part $h$ does not give any propagating mode (in other words, there is no on-shell pole in the two point function), we can remove it from the interaction term by a field redefinition. 
Indeed, if one performs a transformation,
\begin{align}
\label{redef}
h_{\mu\nu}
\to
h_{\mu\nu}-{4\over3m^2 M}\left(\eta_{\mu\nu}+{2\over m^2}\partial_\mu\partial_\nu\right) F_{\rho\sigma}F^{\rho\sigma},
\end{align}
the Lagrangian is given by
\begin{align}
\Delta\mathcal{L}=&-\frac{1}{4}h^{\mu\nu}\mathcal{E}_{\mu\nu}^{\alpha\beta}h_{\alpha\beta}-\frac{m^2}{8}(h_{\mu\nu}^2-h^2)
-\frac{4}{3m^4 M^2}F_{\rho\sigma} F^{\rho\sigma}(2m^2+\Box)F_{\alpha\beta}F^{\alpha\beta}\, .
\end{align}
The BH entropy in this model is smaller than that of the Einstein-Maxwell theory due to the negative coefficient of the $F^4$ term. Since the BH entropy is invariant under field redefinition, the same conclusion applies to the original Lagrangian~\eqref{counterexample}.
This is a simple example which contains more UV degrees of freedom, but gives a smaller BH entropy compared to the Einstein-Maxwell theory.

In the language of \cite{Cheung:2018cwt}, this loophole comes from the fact that the Euclidean action corresponding to the solution of the equation of motion need not be a local minimum with respect to the auxiliary component $h$. We also note that~\cite{Cheung:2018cwt} made an assumption on the UV theory that its Euclidean action with a vanishing UV field $\chi=0$ is equivalent to that of the Einstein-Maxwell theory for any configuration of the metric and the gauge field. However, this assumption is not invariant under field redefinition, like Eq.~\eqref{redef}.

In our argument, on the other hand, the model~\eqref{counterexample} is excluded by requiring a mild UV behavior of scattering amplitudes at large $s$. 
Also, our argument based on scattering amplitudes does not suffer from ambiguity associated with field redefinition.


\begin{thebibliography}{99}

\bibitem{Vafa:2005ui} 
  C.~Vafa,
  hep-th/0509212.

\bibitem{Brennan:2017rbf} 
  T.~D.~Brennan, F.~Carta and C.~Vafa,
  arXiv:1711.00864 [hep-th].

\bibitem{ArkaniHamed:2006dz} 
  N.~Arkani-Hamed, L.~Motl, A.~Nicolis and C.~Vafa,
  JHEP {\bf 0706}, 060 (2007)
  doi:10.1088/1126-6708/2007/06/060
  [hep-th/0601001].

\bibitem{Nakayama:2015hga} 
  Y.~Nakayama and Y.~Nomura,
  Phys.\ Rev.\ D {\bf 92}, no. 12, 126006 (2015)
  doi:10.1103/PhysRevD.92.126006
  [arXiv:1509.01647 [hep-th]].

\bibitem{Harlow:2015lma} 
  D.~Harlow,
  JHEP {\bf 1601}, 122 (2016)
  doi:10.1007/JHEP01(2016)122
  [arXiv:1510.07911 [hep-th]].

\bibitem{Benjamin:2016fhe} 
  N.~Benjamin, E.~Dyer, A.~L.~Fitzpatrick and S.~Kachru,
  JHEP {\bf 1608}, 041 (2016)
  doi:10.1007/JHEP08(2016)041
  [arXiv:1603.09745 [hep-th]].

\bibitem{Montero:2016tif} 
  M.~Montero, G.~Shiu and P.~Soler,
  JHEP {\bf 1610}, 159 (2016)
  doi:10.1007/JHEP10(2016)159
  [arXiv:1606.08438 [hep-th]].
  
\bibitem{Horowitz:2016ezu} 
  G.~T.~Horowitz, J.~E.~Santos and B.~Way,
  Class.\ Quant.\ Grav.\  {\bf 33}, no. 19, 195007 (2016)
  doi:10.1088/0264-9381/33/19/195007
  [arXiv:1604.06465 [hep-th]].

\bibitem{Cottrell:2016bty} 
  G.~Shiu, P.~Soler and W.~Cottrell,
  arXiv:1611.06270 [hep-th].

\bibitem{Crisford:2017gsb} 
  T.~Crisford, G.~T.~Horowitz and J.~E.~Santos,
  Phys.\ Rev.\ D {\bf 97}, no. 6, 066005 (2018)
  doi:10.1103/PhysRevD.97.066005
  [arXiv:1709.07880 [hep-th]].
  
\bibitem{Yu:2018eqq} 
  T.~Y.~Yu and W.~Y.~Wen,
  Phys.\ Lett.\ B {\bf 781}, 713 (2018)
  doi:10.1016/j.physletb.2018.04.060
  [arXiv:1803.07916 [gr-qc]].

\bibitem{Hebecker:2017uix} 
  A.~Hebecker and P.~Soler,
  JHEP {\bf 1709}, 036 (2017)
  doi:10.1007/JHEP09(2017)036
  [arXiv:1702.06130 [hep-th]].
  
\bibitem{Cheung:2018cwt} 
  C.~Cheung, J.~Liu and G.~N.~Remmen,
  arXiv:1801.08546 [hep-th].

\bibitem{Brown:2015iha} 
  J.~Brown, W.~Cottrell, G.~Shiu and P.~Soler,
  JHEP {\bf 1510}, 023 (2015)
  doi:10.1007/JHEP10(2015)023
  [arXiv:1503.04783 [hep-th]].

\bibitem{Brown:2015lia} 
  J.~Brown, W.~Cottrell, G.~Shiu and P.~Soler,
  JHEP {\bf 1604}, 017 (2016)
  doi:10.1007/JHEP04(2016)017
  [arXiv:1504.00659 [hep-th]].
  
\bibitem{Heidenreich:2015nta} 
  B.~Heidenreich, M.~Reece and T.~Rudelius,
  JHEP {\bf 1602}, 140 (2016)
  doi:10.1007/JHEP02(2016)140
  [arXiv:1509.06374 [hep-th]].

\bibitem{Heidenreich:2016aqi} 
  B.~Heidenreich, M.~Reece and T.~Rudelius,
  JHEP {\bf 1708}, 025 (2017)
  doi:10.1007/JHEP08(2017)025
  [arXiv:1606.08437 [hep-th]].

\bibitem{Lee:2018urn} 
  S.~J.~Lee, W.~Lerche and T.~Weigand,
  arXiv:1808.05958 [hep-th].

\bibitem{Cheung:2014ega} 
  C.~Cheung and G.~N.~Remmen,
  JHEP {\bf 1412}, 087 (2014)
  doi:10.1007/JHEP12(2014)087
  [arXiv:1407.7865 [hep-th]].

\bibitem{Andriolo:2018lvp} 
  S.~Andriolo, D.~Junghans, T.~Noumi and G.~Shiu,
  Fortsch.\ Phys.\  {\bf 66}, no. 5, 1800020 (2018)
  doi:10.1002/prop.201800020
  [arXiv:1802.04287 [hep-th]].

\bibitem{Kats:2006xp} 
  Y.~Kats, L.~Motl and M.~Padi,
  JHEP {\bf 0712}, 068 (2007)
  doi:10.1088/1126-6708/2007/12/068
  [hep-th/0606100].

\bibitem{Adams:2006sv} 
  A.~Adams, N.~Arkani-Hamed, S.~Dubovsky, A.~Nicolis and R.~Rattazzi,
  JHEP {\bf 0610}, 014 (2006)
  doi:10.1088/1126-6708/2006/10/014
  [hep-th/0602178].

\bibitem{Bellazzini:2016xrt} 
  B.~Bellazzini,
  JHEP {\bf 1702}, 034 (2017)
  doi:10.1007/JHEP02(2017)034
  [arXiv:1605.06111 [hep-th]].
  

\bibitem{Camanho:2014apa} 
  X.~O.~Camanho, J.~D.~Edelstein, J.~Maldacena and A.~Zhiboedov,
  JHEP {\bf 1602}, 020 (2016)
  doi:10.1007/JHEP02(2016)020
  [arXiv:1407.5597 [hep-th]].

\bibitem{Li:2017lmh} 
  D.~Li, D.~Meltzer and D.~Poland,
  JHEP {\bf 1712}, 013 (2017)
  doi:10.1007/JHEP12(2017)013
  [arXiv:1705.03453 [hep-th]].

\bibitem{Afkhami-Jeddi:2018own} 
  N.~Afkhami-Jeddi, S.~Kundu and A.~Tajdini,
  arXiv:1805.07393 [hep-th].

\bibitem{Goon:2016une} 
  G.~Goon and K.~Hinterbichler,
  JHEP {\bf 1702}, 134 (2017)
  doi:10.1007/JHEP02(2017)134
  [arXiv:1609.00723 [hep-th]].
  
\bibitem{Elvang:2013cua} 
  H.~Elvang and Y.~t.~Huang,
  arXiv:1308.1697 [hep-th].

\bibitem{Gross:1985rr} 
  D.~J.~Gross, J.~A.~Harvey, E.~J.~Martinec and R.~Rohm,
  Nucl.\ Phys.\ B {\bf 267}, 75 (1986).
  doi:10.1016/0550-3213(86)90146-X

\bibitem{AlvarezGaume:1986jb} 
  L.~Alvarez-Gaume, P.~H.~Ginsparg, G.~W.~Moore and C.~Vafa,
  Phys.\ Lett.\ B {\bf 171}, 155 (1986).
  doi:10.1016/0370-2693(86)91524-8

\bibitem{Ooguri:2016pdq} 
  H.~Ooguri and C.~Vafa,
  Adv.\ Theor.\ Math.\ Phys.\  {\bf 21}, 1787 (2017)
  doi:10.4310/ATMP.2017.v21.n7.a8
  [arXiv:1610.01533 [hep-th]].

\bibitem{Ibanez:2017kvh} 
  L.~E.~Ibanez, V.~Martin-Lozano and I.~Valenzuela,
  JHEP {\bf 1711}, 066 (2017)
  doi:10.1007/JHEP11(2017)066
  [arXiv:1706.05392 [hep-th]].

\bibitem{Hamada:2017yji} 
  Y.~Hamada and G.~Shiu,
  JHEP {\bf 1711}, 043 (2017)
  doi:10.1007/JHEP11(2017)043
  [arXiv:1707.06326 [hep-th]].

\bibitem{Gonzalo:2018tpb} 
  E.~Gonzalo, A.~Herraez and L.~E.~Ibanez,
  JHEP {\bf 1806}, 051 (2018)
  doi:10.1007/JHEP06(2018)051
  [arXiv:1803.08455 [hep-th]].

\bibitem{Gonzalo:2018dxi} 
  E.~Gonzalo and L.~E.~Ibanez,
  Phys.\ Lett.\ B {\bf 786}, 272 (2018)
  doi:10.1016/j.physletb.2018.09.034
  [arXiv:1806.09647 [hep-th]].
  
  
\bibitem{Arkani-Hamed:2017jhn} 
  N.~Arkani-Hamed, T.~C.~Huang and Y.~t.~Huang,
  arXiv:1709.04891 [hep-th].

\bibitem{Afkhami-Jeddi:2016ntf} 
  N.~Afkhami-Jeddi, T.~Hartman, S.~Kundu and A.~Tajdini,
  JHEP {\bf 1712}, 049 (2017)
  doi:10.1007/JHEP12(2017)049
  [arXiv:1610.09378 [hep-th]].

\bibitem{Wald:1993nt} 
  R.~M.~Wald,
  Phys.\ Rev.\ D {\bf 48}, no. 8, R3427 (1993)
  doi:10.1103/PhysRevD.48.R3427
  [gr-qc/9307038].

\bibitem{Cheung:2016wjt} 
  C.~Cheung and G.~N.~Remmen,
  Phys.\ Rev.\ Lett.\  {\bf 118}, no. 5, 051601 (2017)
  doi:10.1103/PhysRevLett.118.051601
  [arXiv:1608.02942 [hep-th]].
  
\bibitem{Jacobson:1993xs} 
  T.~Jacobson and R.~C.~Myers,
  Phys.\ Rev.\ Lett.\  {\bf 70}, 3684 (1993)
  doi:10.1103/PhysRevLett.70.3684
  [hep-th/9305016].

\bibitem{Liko:2007vi} 
  T.~Liko,
  Phys.\ Rev.\ D {\bf 77}, 064004 (2008)
  doi:10.1103/PhysRevD.77.064004
  [arXiv:0705.1518 [gr-qc]].

\bibitem{Sarkar:2010xp} 
  S.~Sarkar and A.~C.~Wall,
  Phys.\ Rev.\ D {\bf 83}, 124048 (2011)
  doi:10.1103/PhysRevD.83.124048
  [arXiv:1011.4988 [gr-qc]].
  
\bibitem{deRham:2014zqa} 
  C.~de Rham,
  Living Rev.\ Rel.\  {\bf 17}, 7 (2014)
  doi:10.12942/lrr-2014-7
  [arXiv:1401.4173 [hep-th]].
  
  \end{thebibliography}
\end{document}